\documentclass[conference]{IEEEtran}
\ifCLASSINFOpdf
   \usepackage[pdftex]{graphicx}
  \graphicspath{{../pdf/}{../jpeg/}}
\else
\fi
\usepackage{amsmath}
\usepackage{listings}

\usepackage[T1]{fontenc}
\usepackage{multirow}
\usepackage{suffix}
\usepackage{url}
 \usepackage{acronym}
\usepackage{booktabs}
\usepackage{empheq}
\usepackage{xcolor}
\usepackage{enumitem}
\usepackage{threeparttable}
\usepackage{paralist}
\usepackage{array}

\usepackage{cite}
\usepackage[numbers]{natbib}     

\usepackage{expl3}
\ExplSyntaxOn
\newcommand\latinabbrev[1]{
  \peek_meaning:NTF . {
    #1\@}%
  { \peek_catcode:NTF a {
      #1., \@ }%
    {#1., \@}}}
\ExplSyntaxOff

\definecolor{lightpurple}{rgb}{0.8,0.8,1}
\definecolor{codebg}{RGB}{255,255,255}
\definecolor{commentcolor}{RGB}{11,140,11}
\newsavebox{\supbox}
\newcommand{\bsup}{\begin{lrbox}{\supbox}$\tt\scriptstyle}
\newcommand{\esup}{$\end{lrbox}{}^{\usebox{\supbox}}}
\def\eg{\latinabbrev{e.g}}
\def\ie{\latinabbrev{i.e}}

\newcolumntype{L}[1]{>{\raggedright\let\newline\\\arraybackslash\hspace{0pt}}m{#1}}
\newcolumntype{C}[1]{>{\centering\let\newline\\\arraybackslash\hspace{0pt}}m{#1}}
\newcolumntype{R}[1]{>{\raggedleft\let\newline\\\arraybackslash\hspace{0pt}}m{#1}}

\begin{document}
%
\title{An IDE-Based Context-Aware Meta Search Engine}


\author{\IEEEauthorblockN{Mohammad Masudur Rahman, Shamima Yeasmin,  Chanchal K. Roy }
\IEEEauthorblockA{Department of Computer Science\\
University of Saskatchewan, Canada\\
\{mor543, shy942, ckr353\}@mail.usask.ca}
}



%


\maketitle


\begin{abstract}
Traditional web search forces the developers to leave their working environments and look for solutions in the web browsers. It often does not consider the context of their programming problems. The context-switching between the web browser and the working environment is time-consuming and distracting, and the keyword-based traditional search often does not help much in problem solving. In this paper, we propose an Eclipse IDE-based web search solution that collects the data from three web search APIs-- Google, Yahoo, Bing and a programming Q \& A site-- StackOverflow. It then provides search results within IDE taking not only the content of the selected error into account but also the problem context, popularity and search engine recommendation of the result links. Experiments with 25 runtime errors and exceptions show that the proposed approach outperforms the keyword-based search approaches with a recommendation accuracy of 96\%. We also validate the results with a user study involving five prospective participants where we get a result agreement of 64.28\%. While the preliminary results are promising, the approach needs to be further validated with more errors and exceptions followed by a user study with more participants to establish itself as a complete IDE-based web search solution.

\end{abstract}


%

\IEEEpeerreviewmaketitle

\section{Introduction}
During development and maintenance of a software system, software developers face different programming challenges,  and one of them is runtime error or exception. Eclipse IDE facilitates to diagnose the encountered errors or exceptions and developers get valuable information (\ie\ clues) for fixation from the stack traces produced by the IDE. However, the information from the stack trace alone may not be helpful enough for the fixation, especially when the developers are novice or the encountered problems are relatively unfamiliar. Thus, for more informative and up-to-date solutions, developers are often forced to dig into the world wide web and look for the fixation. In a study by \citet{twostudy}, developers, on average, spent about 19\% of their programming time for surfing the web for information. \citet{codetrail} made a study where they analyzed the events produced by the web browser and the IDE in temporal proximity, and concluded that 23\% web pages visited were related to software development or maintenance.


Finding the working solution to a programming problem in the web is a matter of web surfing as well as programming experience. Novice developers involved in software development or maintenance are often found spending a lot of time to look for such solutions. Traditional web search forces the developers to leave the working environment (\ie\ IDE) and look for the solution in the web browsers. The context-switching between IDE and the web browser is distracting and time-consuming. Moreover, checking relevance from hundreds of search results is a cognitive burden on the novice developers. 


Existing studies focus on integrating commercial-off-the-shelf (COTS) tools into Eclipse IDE \cite{ges}, recommending StackOverflow posts and displaying them within the IDE environment \cite{context}, embedding web browser inside the IDE \cite{embed} and so on. \citet{context} propose an IDE-based recommendation system for runtime exceptions. They extract the question and answer posts from StackOverflow data dump and suggest  posts relevant to the occurred exceptions considering the context from the stack trace information generated by the IDE. They also suggest a nice solution to the context-switching issue through visualization of the solution within the IDE. However, the proposed approach suffers from several limitations.  First, they consider only one source (\eg\ StackOverflow Q \& A site) rather than the whole web for information and thus, their search scope is limited. Second, the developed corpus cannot be easily updated and is subjected to the availability of the data dump. For example, they use the StackOverflow data dump of September 2011, that means it cannot provide help or suggestions to the recently introduced software bugs or errors after September 2011. Third, the visualization of the solutions is not efficient as it uses plain text to show the post contents such as source code, stack trace and discussion. Thus the developers do not really experience the style and presentation of a web page.


In this paper, we propose an Eclipse IDE-based search solution called \emph{SurfClipse} to the encountered errors or exceptions  which addresses the concerns identified in case of existing approaches. We package the solution as an Eclipse plug-in  which (1) exploits the search and ranking algorithms of three reliable web search engines (\eg\ Google, Bing and Yahoo) and a programming Q \& A site (\eg\ StackOverflow) through their API endpoints, (2) provides a content (\eg\ error message), context (\eg\ stack trace and surrounding source code of the subject error), popularity and search engine recommendation (of result links) based filtration and ranking on the extracted results of step one, (3) facilitates the most recent solutions, accesses the complete and extensible solution set and pulls solutions from numerous forums, discussion boards, blogs, programming Q \& A sites and so on, and (4) provides a real web surfing experiences within the IDE context using Java based browser.

We conduct an experiment on \emph{SurfClipse} with 25 programming errors and exceptions which shows interesting findings. Our approach recommends correct solutions for 24 errors and exceptions, which provides an accuracy of 96\%, and most of the solutions are provided within the top five results. In order to validate the applicability of the proposed approach, we conduct a user study involving five prospective participants. We experience 64.28\% agreement between the solutions chosen by the participants and the solutions proposed by our approach. Given that relevance checking of a solution to a programming problem is a subjective process and controlled by different subjective factors, our approach performs considerably well. While the preliminary results are promising, the proposed approach needs to be further validated with more errors and exceptions followed by a user study with more users to establish itself as a complete IDE-based web search solution.

\section{Motivation}
\label{sec:motivation}
\begin{figure*}[!t]

\centering
\includegraphics[scale=.40]{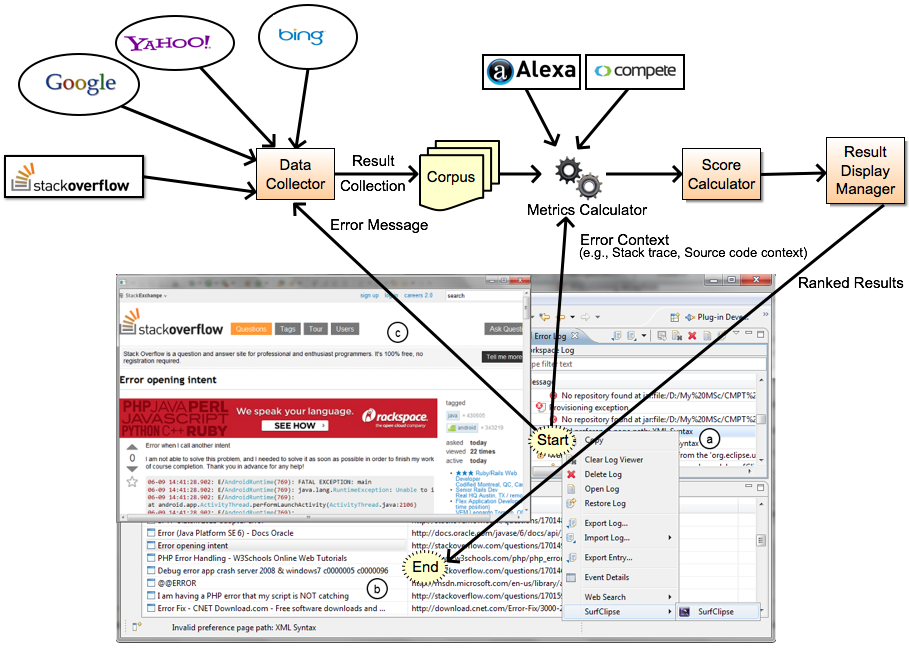}
\caption{Schematic Diagram of the Proposed Approach}
\label{fig:sysdiag}
\vspace{-0.4cm}
\end{figure*}

Traditional web search forces the developers to leave the working environment (\ie\ IDE) and look for the solution in the web browsers. In contrast, if the developer chooses \emph{SurfClipse}, it allows to check the search results within the context of IDE (\eg\ Fig. \ref{fig:sysdiag}-(b)). Once she selects an error message using context menu option (\eg\ Fig. \ref{fig:sysdiag}-(a)), the plug-in pulls results from three reliable search engines and one programming Q \& A site against that error message. Then, it calculates the proposed metrics for each result related to the error content, error context, popularity and search engine recommendation to determine its relevance with the occurred error or exception, and then sorts and displays the results. Moreover, the plug-in allows the developer to browse the solution on a Java-based web browser (\eg\ Fig. \ref{fig:sysdiag}-(c)) without leaving the context of the IDE which makes it time-efficient and flexible to use. The plug-in by \citet{context} also shows the results within the context of the IDE; however, (1) the result set is limited (\ie\ only from StackOverflow and does not consider the whole web), (2) cannot address newly introduced issues (\ie\ fixed corpus and subjected to the availability of StackOverflow data dump), (3) only considers stack trace information as problem context, and (4) the developer cannot enjoy the web browsing experience.

\section{Proposed Approach}
\label{sec:search}
Fig. \ref{fig:sysdiag} shows the schematic diagram of our proposed approach for IDE-based web search. Once the developer selects an exception from the \emph{Error Log} or \emph{Console View} of Eclipse IDE, our approach collects necessary information about it such as error message, stack trace and source code context. Then, it collects the results from three reliable search engines (\eg\ Google, Bing and Yahoo) and one programming Q \& A site (\eg\ StackOverflow) through API endpoints against the error message and develops the corpus. The proposed approach then considers the context of the occurred error or exception, popularity and search engine recommendation of the collected results and calculates the proposed metrics to determine their acceptability and relevance with the target exception. Once the final scores are calculated from those metrics, the results are filtered, sorted and displayed to the developers within the context of IDE. The following sections discuss about the proposed metrics and the scores we use in our approach.


\subsection{Proposed Metrics}
\label{sec:metrics}

\subsubsection{Search Engine Weight Based Score ($S_{sew}$)}
According to \emph{Alexa}\footnote{http://www.alexa.com/topsites, Visited on June, 2013}, one of the widely recognized web traffic data providers, Google ranks second, Yahoo ranks fourth and Bing ranks sixteenth among all websites in the web this year. While these ranks indicate their popularity (\eg\ site traffic) and reliability (\ie\ users' trust) as information service providers,  it is reasonable to think that search results from different search engines of different ranks have different levels of acceptance. We conduct an experiment with 75 programming task and exception related queries\footnote{http://homepage.usask.ca/~mor543/surfclipse/query.txt} against those search engines and a programming Q \& A site (\eg\ StackOverflow) to determine the relative weights or acceptance. We collect the top 15 search results for each query from each search tool and get their \emph{Alexa ranks}. Then, we consider the \emph{Alexa ranks} of all result links provided by each search tool and calculate the average rank for a result link provided by them. The average rank for each search tool is then normalized and inversed which provides a value between 0 and 1. We get a normalized weight of 0.41 for Google, 0.30 for Bing, 0.29 for Yahoo and 1.00 for StackOverflow. The idea is that if a result link against a single query is found in all three search engines, it gets the search engine scores (\ie\ confidence) from all three of them which sum to 1.00. StackOverflow has drawn the attention of a vast community (1.7 million\footnote{http://en.wikipedia.org/wiki/Stackoverflow, Visited on June, 2013}) of programmers and software professionals, and it also has a far better \emph{average Alexa rank} than that of the search engines; therefore, the results returned from StackOverflow are provided a search engine score (\ie\ confidence) of 1.00.



\subsubsection{Title Matching Score ($S_{title}$)}
During errors or exceptions, the IDE or Java framework generally issues  notifications from a fixed set of error or exception messages. Thus, there is a great chance that a result page titled with an error or exception message similar to the search query would discuss about the encountered problem by the developer and hopefully would contain relevant information for fixation. We consider the cosine similarity\footnote{http://en.wikipedia.org/wiki/Cosine\_similarity, Visited on June, 2013} measure between search query and the result title as \emph{Title Matching Score}, $S_{title}$ which provides a value between zero and one. Here, zero indicates complete dissimilarity and one indicates complete similarity between search query and the title of the result. 

\subsubsection{Stack Trace Matching Score ($S_{st}$)}
To solve the programming errors or exceptions, the associated contextual information such as stack trace generated by the IDE plays an important role. Stack trace contains the error or exception type, system messages and method call references in different source files. In this research, we consider an incentive to the result links containing stack traces similar to that of the selected error or exception. The result links may contain stack traces; however, they are likely to be generated for different contexts or different user programs. Thus, the complete lexical similarity between them and the target stack trace is not likely and partial similarity is a suitable choice to determine their relevance. \emph{SimHash Algorithm} performs better for partial similarity matching between two blocks of contents \cite{simhash} and we use it to determine relevance between corresponding stack traces. We extract the stack trace information from the result page through HTML scrapping and apply \emph{SimHash Algorithm} on both stack traces. We get their \emph{SimHash} values and determine the \emph{Hamming distance}. We repeat the process for all result links containing stack traces and use equation \eqref{eq:stscore} to determine their \emph{Stack Trace Scores}.
\begin{equation}\label{eq:stscore}
S_{st}=1-\frac{{d_{k}}-\alpha}{max(d_{k})-\alpha}
\end{equation}
Here, $d_{k}$ represents the \emph{Hamming distance} between the \emph{Hash values} of each result stack trace and the stack trace of the selected exception, $max(d_{k})$ represents the maximum \emph{Hamming distance} found, $\alpha$ represents the minimum \emph{Hamming distance} found and $S_{st}$ refers to the \emph{Stack Trace Score}. 
The score values from zero to one and indicates the relevance of result link with the target exception in terms of stack trace information.


\subsubsection{Source Code Context Matching Score ($S_{cc}$)}
Sometimes, stack trace may not be enough for problem fixation and developers post related source code in forums and discussion boards for clarification. We are interested to check if the source code contexts of the discussed errors or exceptions in the result links are similar to that of the selected exception from IDE. The code contextual similarity is possible; because, the developers often reuse code snippets from programming Q \& A sites, forums or discussion boards in their program directly or with minor modifications. Therefore, a result link containing source code snippet similar to the surrounding code block of the selected error or exception location is likely to discuss relevant issues that the developer needs. We consider three lines before and after the affected line in the source file as the source code context of the error or exception and extract the code snippets from result links though HTML scrapping. Then, we apply \emph{SimHash Algorithm} on both code contexts and generate their \emph{SimHash values}. We use equation \eqref{eq:stscore} to determine \emph{Source Code Context Matching Score} for each result link. The score values from zero to one and it indicates the relevance of the result link with the target error in terms of the context of source  code.

\subsubsection{StackOverflow Vote Score ($S_{so}$)}
StackOverflow Q \& A site maintains score for each asked question post, answer post and comment, and the score can be considered as a social and technical recognition of their merit \cite{so}. Here, the user can up-vote a post if he/she likes something about it, and can also down-vote if the post content seems incomplete, confusing or not helpful. Thus, the difference between up-vote and down-vote, also called the score of a post, is considered as an important metric for evaluation of the quality of the question or solution posted. In our research, we consider this score for the result links from StackOverflow site. Once the corpus is formed dynamically, we consider the scores of all StackOverflow result links and calculate their normalized scores using equation \eqref{eq:soscore}.
\vspace{-1mm}
\begin{equation}\label{eq:soscore}
S_{so}=\frac{{SO_{k}}-\beta}{max(SO_{k})-\beta}
\end{equation}
\vspace{1mm}
Here, $SO_{k}$ refers to the StackOverflow post score, $max(SO_{k})$ represents the maximum score found, $\beta$ represents the minimum post score found and $S_{so}$ is the \emph{StackOverflow Vote Score} for the result link. The score values from zero (\ie\ least significant) to one (\ie\ most significant) and it indicates the subjective evaluation of the link by a large developer crowd of 1.7 million.

\subsubsection{Top Ten Score ($S_{tt}$)}
Ranking of first 10 results  is considered very important for any search engine. Generally, users look for the solution in the first 10 results before switching to next query. In this research, we are interested to exploit the top ten ranking information provided by all web search APIs and StackOverflow, and we provide incentives to the result links found in top 10 positions. We provide a normalized score using equation \eqref{eq:top10} for each top result.
\begin{equation}\label{eq:top10}
S_{tt}=1-\frac{\bar{P_{k}}-1}{10}
\end{equation}
Here, $\bar{P_{k}}$ represents the average position of a result link in top 10 ranking of each search tool and $S_{TT}$ represents the \emph{Top Ten Score}.

\subsubsection{Page Rank Score ($S_{pr}$)}
Page Rank score is considered as an interesting metric to determine the relative importance of a list of web pages or web sites based on their interconnectivity. The idea is that a page sets hyper links to another page or site when its contents are somehow related  and the first page recommends browsing the other page or site to the visitors. This type of recommendation is important as it carries subjective evaluation of the web site by the users  and we are interested to consider that in this research. We calculate \emph{Page Rank Score} as a measure of the worth for recommendation. We develop an interconnected network for all the result entries in the corpus considering their incoming and outgoing links and calculate the score using  \emph{PageRank Algorithm} \cite{pagerank}. The score is also normalized and it values between zero to one.

\subsubsection{Search Traffic Rank Score ($S_{str}$)}
The amount of search traffic to a site can be considered as an important indicator of its popularity. In this research, we consider the relative popularity of the result links found in the corpus. We use the statistical data from two popular site traffic control companies- \emph{Alexa} and \emph{Compete} through their provided APIs and get the average ranking for each result link. Then, based on their ranks, we provide a normalized \emph{Search Traffic Rank Score} between zero and one considering minimum and maximum search traffic ranks found.

\subsection{Result Scores Calculation}
\label{sec:scores}
The proposed metrics focus on four aspects of evaluation for each result entry against the search query. They are-- popularity of the result link, error content-based similarity, error context-based similarity,  and search engine factors. \emph{StackOverflow Vote Score}, \emph{Page Rank Score} and \emph{Search Traffic Rank Score} are considered as the measures of popularity of the result link from different viewpoints. For example, \emph{StackOverflow Vote Score} is directly computed from the votes provided by StackOverflow community, \emph{Search Traffic Rank Score} is calculated from the site ranks provided by \emph{Alexa} and \emph{Compete}, and \emph{Page Rank Score} is based on the interconnectivity among the result links. We consider the average of these component scores as the \emph{Popularity Score}, $S_{pop}$, of the result and use equation \eqref{eq:popularity} to get the score.
\begin{equation}\label{eq:popularity}
S_{pop}=\frac{1}{3}\sum_{t\in\{so,str,pr\}}S_{t}
\end{equation}
\emph{Title Matching Score} measures the content similarity between search query and result title. \emph{Stack Trace Matching Score} and \emph{Source Code Context Matching Score} determine the relevance of the result link based on its contextual similarity with that of the selected error or exception; therefore, they constitute the \emph{Context Relevance Score}, $S_{cxt}$. We get this score using equation \eqref{eq:contextscore}.
\begin{equation}\label{eq:contextscore}
S_{cxt}=\frac{1}{2}\sum_{t \in \{st,cc\}}S_{t}
\end{equation}
\emph{Search Engine Weight Based Score} denotes the relative acceptance of the result link based on its availability in the result sets provided by different search tools, and \emph{Top Ten Score} refers to the relevance of the result link based on its availability in top 10 positions. Thus, \emph{Search Engine Recommendation score}, $S_{ser}$, for the result link can be considered as the product of its acceptance (\ie\ confidence) measure and relevance measure. We use equation \eqref{eq:recom} to get the score.
\begin{equation}\label{eq:recom}
S_{ser}=\prod_{t\in\{sew,tt\}}S_{t}
\end{equation}


\section{Experimental Results and Validation}
\label{sec:results}

\subsection{Experimental Results}
In our experiment, we select 25 runtime errors and exceptions related to Eclipse plug-in development, and collect associated information such as error or exception messages, stack traces and source code context. We then use those information (\eg\ error content and context) to search for solution using our approach. We also perform extensive web search manually with different available search engines and find out the solutions for all errors and exceptions. We should  note that we choose the most appropriate solution as the accepted one for each exception or error 

During search and ranking, we consider different combinations of the calculated scores (Section \ref{sec:scores}) to get the search results for each query, and then identify the accepted solution within the top 10 and top 20 result entries. We also calculate the average ranks of the identified solutions. Table \ref{table:results} shows the results of our conducted experiments. We note that the approach that considers the encountered error context, popularity and search engine recommendation about the result links in addition to the query error message for relevance and acceptance (\ie\ confidence) checking, outperforms the traditional keyword-based search (\ie\ that only considers the keywords in the query) both in terms of recommendation accuracy and average rankings of the solutions.

\begin{table}[!t]
\caption{Search Results against 25 Queries}
\vspace{-0.3cm}
\label{table:results}
\centering
\begin{threeparttable}[b]
\begin{tabular}{|l|c|c|c|c|}
\hline
\textbf{Scores} & $\mathbf{Soln_{10}}$\tnote{1} & $\mathbf{R_{10}}$\tnote{2} & $\mathbf{Soln_{20}}$\tnote{3} & $\mathbf{R_{20}}$\tnote{4}\\
\hline
$S_{cnt}$ (keyword based) & 10 & 3.6000 & 16 & 8.6250 \\
\hline
$S_{cnt},S_{cxt}$ & 11 & 3.0000 & 16 & 7.4375\\
\hline
$S_{cnt},S_{pop}$ & 13 & 4.6920 & 18 & 8.1111\\
\hline
$S_{cnt},S_{ser}$ & 23 & 4.3910 & 23 & 4.3913\\
\hline
$S_{cnt},S_{cxt},S_{pop}$ & 13 & 4.0769 & 18 & 7.6111\\
\hline
$S_{cnt},S_{cxt},S_{ser}$ & \textbf{24} & \textbf{4.4583} & \textbf{24} & \textbf{4.4583}\\
\hline
$S_{cnt},S_{cxt},S_{pop},S_{ser}$ & 23 & 4.2608 & \textbf{24} & \textbf{4.5416}\\
\hline
\end{tabular}
\begin{tablenotes}
   	\item[1] Solutions found within first 10 results.
	\item[2] Average rank for solutions within first 10 results.
	\item[3] Solutions found within first 20 results.
	\item[4] Average rank for solutions within first 20 results.
 \end{tablenotes}
\end{threeparttable}
\vspace{-.2cm}
\end{table}

\subsection{Validation of the Proposed Approach by User Study}
We select five frequent exceptions from the list used in the experiment and involve five graduate research students in the user study. Each exception was associated with ten solutions recommended by our approach and the participants were instructed to choose one or more solutions for each exception. We should note that the answers were not ranked, and were presented in a random order to the participants. The idea is to prevent the bias of the participants of selecting top answers, and to discover the subjective views about the relevance of solutions against an exception. We determine the matching between the solutions chosen by the participants and the top five results recommended by our approach. Table \ref{table:ustudy} shows the results of the conducted user study. We get 64.28\% agreement between the responses of the participants and our proposed approach. Given that relevance checking of a solution against the selected error is a completely subjective process and controlled by various subjective factors, the agreement amount is a significant one.
\begin{table}[!t]
\caption{Results from User Study}
\vspace{-0.3cm}
\label{table:ustudy}
\centering
\begin{threeparttable}[b]
\begin{tabular}{|l|c|c|c|}
\hline
\textbf{Question ID} & \textbf{ANSR}\tnote{1} & \textbf{ANSM}\tnote{2} & \textbf{Agreement}\tnote{3}\\
\hline
$Q_1$ & 2.8 & 2.0 &  71.43\% \\
\hline
$Q_2$ & 4.6 & 2.8 &  60.87\%\\
\hline
$Q_3$ & 4.6 & 2.4 &  52.17\%  \\
\hline
$Q_4$ & 4.2 & 3.0 &  71.43\% \\
\hline
$Q_5$ & 5.8 & 3.8 &  65.52\% \\
\hline
\textbf{Overall} & 4.4 & 2.8 & \textbf{64.28\%}\\
\hline
\end{tabular}
\begin{tablenotes}
   	\item[1] Avg. no. of solutions recommended by participants 
	\item[2] Avg. no. of solutions matched with that by proposed approach
	\item[3] Percentage of agreement between solutions
  \end{tablenotes}
\end{threeparttable}
\vspace{-.3cm}
\end{table}

\section{Related Works}
\label{sec:threats}
Existing studies related to our research focus on integrating commercial-off-the-shelf (COTS) tools into Eclipse IDE \cite{ges}, recommending StackOverflow posts and displaying within IDE environment \cite{context, seahawk}, embedding web browser inside the IDE \cite{embed} for code example recommendation and so on. In this paper, we propose a novel approach that exploits result data from the state of art web search APIs and provides filtered and ranked search results taking problem content, context, result link's popularity and search engine recommendation about the result links into consideration. Our proposed approach not only collects solution posts from a large set of forums, discussion boards and Q \& A sites with the help of search enginies, but also ensures the access to the most recent content of StackOverflow through API access. However, the existing approaches by \citet{context} and \citet{seahawk} provide results from a single and fixed sized data dump of StackOverflow and therefore, the results do not contain the most recent posts (\ie\ discussing the most recent errors or exceptions) from StackOverflow as well as the promising solutions from other programming Q \& A sites.

\section{Conclusion and Future Works}
\label{sec:conclusion}
To summarize, we propose a novel IDE-based web search solution that (1) exploits the search and ranking capabilities of three reliable search engines and a programming Q \& A site through their API endpoints, (2) considers not only the content of the search (i.e., query keywords) but also the problem context such as stack trace and source code context, link popularity and link recommendation from the search engines, and (3) provides search result within the context of IDE with web browsing capabilities. We conduct an experiment with 25 runtime errors and exceptions related to Eclipse plug-in development. Our approach recommended solutions with 96\% accuracy which necessarily outperforms the traditional keyword-based search. In order to validate the results, we conduct a user study involving five prospective participants which gave a response agreement of 64.28\%. Given that the relevance checking of a solution against the selected error is  completely a subjective process, the preliminary results are promising. However, the proposed approach needs to be further validated with more errors and exceptions followed by an extensive user study to establish itself as a complete IDE-based web search solution. We also have plans to enable multiprocessing for the application and host it as a web service API so that others can readily use it with real time experience and also can use the API in their own IDEs rather than Eclipse.

\bibliographystyle{plainnat}
\footnotesize
\bibliography{mybib}

\end{document}